\documentclass{pasa}

\jyear{2016}

\title[The ASP Catalogue]{An All-Sky Portable (ASP) Optical Catalogue}
\author[E. Flesch]{\textbf{Eric Wim Flesch }$^{A,B}$\\
\\
\affil{$^A$PO Box 5, Whakatane  3158, New Zealand}
\affil{$^B$Email: eric@flesch.org}}

\begin{document}

\begin{abstract}
This optical catalogue combines the all-sky USNO-B1.0/A1.0 and most-sky APM catalogues, plus overlays of SDSS optical data, into a single all-sky map presented in a sparse binary format which is easily downloaded at 9Gb zipped.  Total count is 1\,163\,237\,190 sources and each has J2000 astrometry, red \& blue magnitudes with PSFs and variability indicator, and flags for proper motion, epoch, and source survey \& catalogue for each of the photometry and astrometry.  The catalogue is available on the PASA datastore at http://dx.doi.org/10.4225/50/5807fbc12595f . 
\end{abstract}


\maketitle

\section{Introduction}

The largest astronomical catalogues are the all-sky optical ones which reach the 20$^{th}$ magnitude.  The only such catalogues are the United States Naval Observatory USNO-B1.0 \cite{USNOB} and its USNO-A \cite{USNOA} antecedents, the SuperCOSMOS Sky Surveys\footnote{The SuperCOSMOS websites are at http://www-wfau.roe.ac.uk/sss/index.html (surveys) and http://ssa.roe.ac.uk (archive)}, and now also the \textsl{Gaia}\footnote{http://www.cosmos.esa.int/web/gaia} DR1 \cite{GAIA} which gives single-band coverage of unprecedented astrometric accuracy.  Also notable are the Automated Plate Measuring (APM) machine catalogue \cite{APM} which covers the sky excluding the Galactic plane, and of course the Sloan Digital Sky Survey (SDSS)\footnote{http://sdss.org} which gives unprecedentedly deep coverage to a third of the sky.  Those catalogues are so large that downloading is foreclosed for most, and distribution usually consists of getting a copy from someone who already has it.

The need is to transcend this barrier and make these data available via standard internet download.  Herewith I present a concise compilation of USNO-B1.0/A1.0, APM and SDSS data which endeavours to do that.  This ``All-Sky Portable'' catalogue (\textit{hereinafter}: ASP) amounts to 1\,163\,237\,190 optical sources over the whole sky, presented in a minimized format of tenth-arcsecond precision astrometry, hundredth-of-magnitude precision red-blue photometry with stellar-fuzzy PSFs, and flags as to proper motion, epoch, variability, and provenance for each of photometry and astrometry.  Total size is 11Gb, zipped to 9Gb.  No \textsl{Gaia} data is used, but photometric comparison is given below in Section 5.  

The concise downloadable size is not ASP's only deliverable.  Also, the presented data is otherwise difficult to obtain in bulk form.  ASP includes all well-detected red/blue sources from the USNO-B1.0/A1.0, APM and SDSS catalogues.  These four catalogues have problematic access in bulk:
\begin{itemize}
	\item The full USNO-B1.0 is 80Gb in size and is not available for download on-line.  Its data is packed into a proprietary format so is readable by dedicated software only.  The USNO website does not provide the bulk catalog on-line for reasons of bandwidth and readability, so it is accessible only via on-line queries.
	\item The APM catalogue is 8Gb in size and was the premier scanned photometric product of its day (turn of the century), but was distributed on tape by individual request only.  Its data are in binary SUN formats with structured headers giving global parameters.  There is no place to download it; it is accessible only via on-line queries.
	\item The SDSS photometric catalogue is 70 terabytes large but has been made available in a condensed form as the ``datasweep'' edition\footnote{at http://data.sdss3.org/sas/dr9/boss/sweeps/dr9} of 300Gb size, beyond the reach of those without large bandwidth.  
	\item The USNO-A1.0 is 6Gb in size and was distributed on 10 CDs in the late 1990's; it presented 1955-epoch POSS-I astrometry not available in its successor USNO-B1.0.  It was twice-superseded and is today not obtainable; it is accessible only via on-line query\footnote{at http://vizier.u-strasbg.fr/viz-bin/VizieR?-source=USNO-A1.0}. 
  \item The SuperCOSMOS catalog (not included in ASP) is 4 terabytes large and is not available for bulk download.
\end{itemize}

Figure 1 shows where ASP presents these catalogues' data on the all-sky map, and the ratios used.  The ASP optical data is useful for efficient all-sky matchings and searches, and for data-driven analysis requiring continual access and which can be used to formulate queries onto more comprehensive on-line catalogues.  Interested users would include those who need rapid and repeated all-sky optical processing, those who require an optical database for a client application, and those with limited bandwidth.  A discussion of the assembly of this data, along with a brief round-up of the original surveys and the USNO-B1.0/A1.0, APM and SDSS catalogues, and details of this catalogue's structure, follows.

\section{Source data and selection}

\begin{figure*}[t] 
\includegraphics[scale=0.425, angle=0]{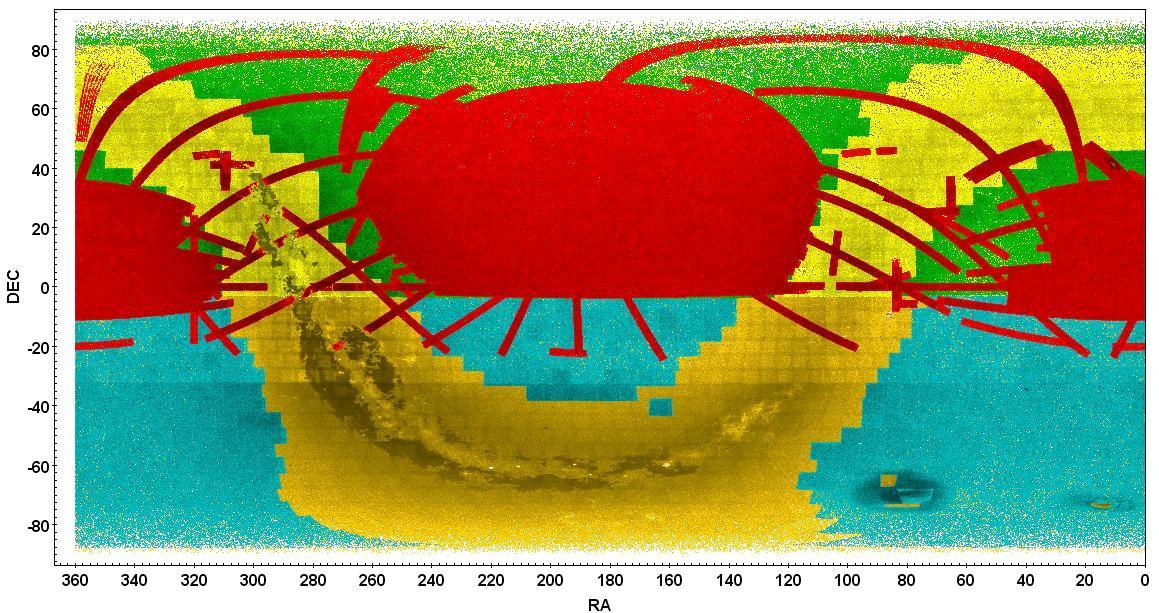} \\
\tiny{Chart produced with TOPCAT \cite{TAYLOR}.}  
\caption{Sky coverage of the ASP catalogue, using 1\% of the data.  Stellar density shows as grey background (compare Figure 3).  Colours designate catalogue/survey photometry coverage as follows:}
\scriptsize
\begin{itemize}
	\item Light Yellow, north ($\delta>-3^{\circ}$) Galactic coverage: USNO-B POSS-I (66\%) and POSS-II (32\%), and USNO-A POSS-I (2\%) sources.  
	\item Medium Yellow, Galactic ($-33^{\circ}<\delta<-3^{\circ}$): USNO-B POSS-I (51\%) and UKST (45\%) sources, and USNO-A POSS-I (4\%) sources.
	\item Dark Yellow, south ($\delta<-33^{\circ}$) Galactic coverage: USNO-B UKST (99\%) and USNO-A UKST (1\%) sources.   
  \item Red: SDSS coverage (52\%) interwoven with coverage mix of background colour (usually green).  
  \item Green ($\delta>-3^{\circ}$): APM POSS-I (38\%), USNO-B POSS-I (25\%) and POSS-II (36\%), and USNO-A POSS-I (1\%) sources.   
  \item Light Blue, ($-33^{\circ}<\delta<-3^{\circ}$): APM UKST (48\%), USNO-B POSS-I (36\%) and UKST (15\%), and USNO-A POSS-I (1\%) sources. 
  \item Blue, ($\delta<-33^{\circ}$): APM UKST (74\%) and USNO-B UKST (26\%) sources.  
\end{itemize}
\end{figure*}

Large sky photographic surveys were done throughout the 2$^{nd}$ half of the 20th century, generally using plates of 6.4$^{\circ}$-square coverage.  Northern sky surveys consisted of the first-epoch National Geographic Society-Palomar Observatory Sky Survey (POSS-I) \cite{POSSI} which took red and blue plates (spaced at 6$^{\circ}$ intervals) on the same night, thus ensuring true red-blue colour even for variable objects, and the second-epoch POSS-II \cite{POSS2} which took deeper plates (spaced at 5$^{\circ}$ intervals) albeit in different epochs.  Southern sky plates were photometrically equivalent to POSS-II; these were taken by United Kingdom Schmidt Telescope surveys (UKST), encompassing SERC / ESO-R / AAO-R projects which covered the entire Southern sky.  

The POSS-I survey used violet \textsl{O} (4050\AA\ centred passband) for blue, while the POSS-II \& UKST surveys used blue-green \textsl{Bj} 4850\AA; red is centred on 6400\AA\ in all those surveys.  In calculating blue-red colour, the POSS-I \textsl{O-E} spread is thus about 1.5 times that of the other surveys' \textsl{Bj-R}.  This greater spread, plus that the POSS-I plates show accurate blue-red colour (due to both plates being taken on the same night, thus minimizing variability issues), makes the POSS-I photometry especially desirable.   Thus, in ASP, it is used as the first choice wherever available which means down to declination $-33^{\circ}$.  The POSS-I and UKST surveys were included in the APM and USNO-B1.0/A1.0 catalogues while the POSS-II survey appears only in the USNO-B1.0\footnote{The full list of USNO-B1.0 surveys is at bottom of http://www.nofs.navy.mil/data/FchPix}.     

The APM catalogue covers sky away from the Galactic plane, see Figure 2 for its coverage.  It consists of 270M optical sources in 959 data files each of which presents one red-blue plate pairing; 498 files present POSS-I data which cover sky north of declination $-3^{\circ}$, and 461 UKST files cover sky south of declination $+3^{\circ}$; thus the equator was doubly covered.  The scanned plates were glass copies which were contrast-enhanced to reveal the faintest sources; this removed any dynamic range but simplified the magnitude calculations.  The APM plate depths were calibrated against modelled sky values, except that the POSS-I \textsl{E} (red) plate depths were \textsl{defined} as equal to 20.0.  Where red and blue sources matched together within 2 arcseconds they were reported as a single two-band source.  There were 4 PSF classes: stellar, fuzzy, blended, and noise.  APM data accepted into ASP were those with at least one band having a stellar or fuzzy PSF, or with both bands showing blended sources which are reported by ASP as either fuzzy or no-PSF.  Single-band POSS-I data can show faint sources but were unfortunately flooded by spurious exposure artefacts \cite{MORX}.  It wasn't possible to distinguish between the valid and invalid single-band POSS-I data, so with much regret I dropped all those unless matched elsewhere; most valid ones are fortunately recovered by the SDSS and USNO-B1.0 POSS-II data.  The APM provides plate astrometry which means 1950's epoch for its POSS-I data.

\begin{figure}[ht] 
\includegraphics[scale=0.2, angle=0]{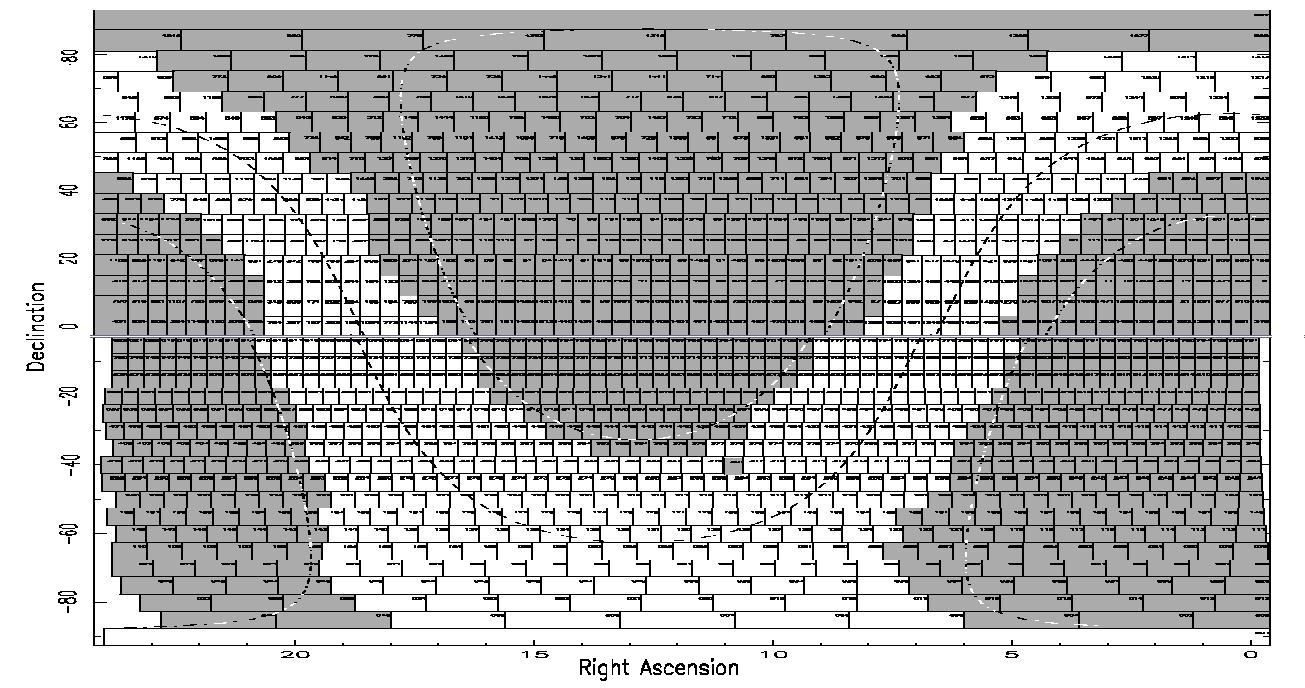} 
\caption{APM sky coverage: dark areas.}   
\end{figure}

The USNO-B1.0 (\textit{hereinafter}: USNO-B) catalogue is all-sky and so covers the Galactic plane exclusively in ASP (except for small numbers from USNO-A1.0).  It presents 1042M optical sources in a processed all-sky map partitioned into 1800 files.  Each record presents red-blue photometry from surveys of two epochs, plus near infrared photometry for the 2$^{nd}$ epoch, as available.  It includes 7435 scanned plates which consist of 937 POSS-I red-blue plate pairings which cover sky of ($-33^{\circ}<\delta<+90^{\circ}$), 897 POSS-II red-blue-infrared plate combos covering sky of ($-3^{\circ}<\delta<+90^{\circ}$), and 822 UKST red-blue-infrared combos covering sky of ($-90^{\circ}<\delta<-3^{\circ}$).  USNO-B identified matching sources within a 3 arcsecond matching radius, but also proper-motion evaluations were done within a 30 arcsecond radius, so enabling recognition of moving stars as a single object.  All astrometry was accordingly calculated into epoch 2000 wherever two epochs of source astrometry were available.  There are 3 PSF classes: stellar, fuzzy, and no-PSF.  USNO-B required two detections (out of 3 bands \& 2 epochs) to accept any source; sources detected on only one plate were dropped. 

The USNO-A1.0/A2.0 catalogues presented astrometry at the plate epoch whereas their successor USNO-B presents astrometry projected to epoch 2000. The USNO-A1.0 presented epoch-1955 POSS-I data for all ($\delta>-33^{\circ}$) whereas the later USNO-A2.0 presented POSS-I data only for ($\delta>-17^{\circ}$), replacing the difference with UKST data.  Therefore the USNO-A1.0 (\textit{hereinafter}: USNO-A) is used by ASP as the complete source for epoch-1955 POSS-I data not already provided by APM, in particular over ($-33^{\circ}<\delta<-3^{\circ}$) and on the Galactic plane; the use of epoch-1955 data is discussed at Section 4 end.  The USNO-A included only sources detected in both red and blue bands within a 2 arcsecond matching radius, and presented magnitudes of tenth-magnitude precision without any PSF information.  ASP uses USNO-A UKST data ($\delta<-33^{\circ}$) only where unmatched to any other.

The SDSS data is taken from their DR9 datasweeps catalogue with 361M optical sources which covers $\approx$35\% of the sky.  The SDSS photometry is 5-band \textsl{ugriz} and the datasweeps acceptance criterion is that each source should have a ``reasonable detection'' in at least one band, i.e. has \textsl{u}$<$22.5 or \textsl{g}$<$22.5 or \textsl{r}$<$22.5 or \textsl{i}$<$22.0 or \textsl{z}$<$21.5 (extinction corrected) (Aihara et al. \shortcite{DR8}, section 5, footnote 84).  Only their red (\textsl{r} 6200\AA) and green (\textsl{g} 4750\AA) bands are presented in ASP; they are AB magnitudes which are \textsl{not} extinction-corrected so can show $\geq$0.3 mag fainter than elsewhere presented.  Many rows (25M) were missing photometry for one or both \textsl{r} \& \textsl{g} bands, and some magnitudes were unexpectedly faint, even into the 30's which could not be useful, especially given that \textsl{r}=24.80 and \textsl{g}=25.11 are the zero-flux magnitudes (Stoughton et al. \shortcite{SDSSEDR}, table 21).  I wished to include only objects with useful \textsl{r} and \textsl{g} magnitudes into ASP, so my simple acceptance criterion is that both bands be brighter than their zero-flux magnitudes, given that the ``reasonable detection'' criterion was already met.  As a supplement, 9M additional SDSS DR8 objects were taken from the 160M-object \textsl{SDSS-XDQSO} quasar targeting catalogue of Bovy et al. \shortcite{XDQSO}; they show extinction-corrected asinh magnitudes (``luptitudes'' in the SDSS lexicon) which meet the ``reasonable detection'' standard and which, for these 9M objects, met my acceptance criterion where the datasweeps AB magnitudes did not.  These look valid on spot checks of SDSS finding charts, and so they are added for completeness.  SDSS photometry provides 2 PSF classes: stellar and galaxy (i.e., fuzzy) which are decided over the whole source bandwidth so are written in ASP as the same for both bands.

\section{Processing and assembly of the ASP catalogue}

Processing commenced with the APM data which present 959 complete red-blue plates which overlap neighbouring plates, thus allowing the use of shared stars to recalibrate the plate depths\footnote{The individual plate calibrations are listed on http://quasars.org/APM-USNOB-plate-calibration.txt}.  The POSS-I (Northern sky) and UKST (Southern sky) plates overlap eachother on the equator, and their red bands are both commensurate with Cousins \textsl{R}; thus they could be calibrated against eachother in bulk.  This comparison showed that, consistent with an earlier exercise \cite{QORG}, the POSS-I \textsl{E} plate depths were listed at 0.16 magnitude too bright as a group.  Thus all APM POSS-I \textsl{E} plate depths and magnitudes were made 0.16 magnitude fainter, and the same done for the POSS-I \textsl{O} magnitudes to preserve the internal APM (\textsl{O-E}) colour. 

Next, red-blue photometry was selected from the USNO-B data, preferring two-band POSS-I magnitudes where available.  The USNO-B data identifies the source survey \& plate for each record; these were used to parcel out its data into 2656 files corresponding to the original plates.  These plate data are incomplete because USNO-B already de-duplicated them across overlapping plates within a 3-arcsec matching radius, but were needed here for calibration purposes.  Tycho-2 stars and other unallocated objects were set aside to be added back in later.  Spurious sources with magnitudes fainter than 23.0 were identified and removed.  USNO-B has 910 plates in common with APM; their photometry were now normalized (calibrated) plate-by-plate$\textstyle^7$ onto their APM equivalents, using shared stars.  The bulk result showed that, in Galactic places not covered by APM, POSS-I magnitudes were to be adjusted by +0.09 in \textsl{E} and by +0.30 in \textsl{O}, and UKST magnitudes were perfectly aligned in \textsl{R} and to be adjusted by +0.40 in \textsl{Bj}; this was done.  USNO-A magnitudes were similarly calibrated\footnote{The individual plate calibrations are listed on http://quasars.org/APM-USNOA-plate-calibration.txt}: the bulk Galactic adjustments for POSS-I magnitudes were +0.1 in both \textsl{E} and \textsl{O}, and for UKST magnitudes were $-0.5$ in \textsl{R} and $-1.1$ in \textsl{Bj}.  

The 910 plates with both APM and USNO-B coverage were now individually combined, matching objects one-to-one across catalogues within a 5.35 arcsec radius; it was important not to generate spurious duplicates, thus the large radius\footnote{Matching radii are found via annulus counts and confirmed by spot checks.  See the Appendix for a discussion of this topic and usage in ASP.}.  The USNO-A POSS-I plate data were now added into the USNO-B/APM plate data in the same way.  Two polar POSS-I plate files were combined into a single polar file; similarly, four POSS-II polar plate files were thus combined.  The outcome was 2700 red-blue plate files corresponding to 936 POSS-I plates, 894 POSS-II plates, and 870 UKST plates.

De-duplication was now done on overlapping same-survey plates, one-to-one to a matching radius of 5.35 arcsec; approx 50 million duplicates were removed.  Numerous spot checks on well-separated duplicates confirmed that duplicates were being accurately targetted; the large radius was needed because astrometric offsets at plate margins can be large, up to 3 arcsec, and the overlapping plates are offset opposite to each other, thus doubling the total offset.  Most duplications came from the APM files which contained full plate data including overlapping areas; also, USNO-B had performed de-duplication of overlapping plates to a 3 arcsecond radius, thus many more were still eligible for de-duplication; similarly for USNO-A.  All these de-duplications were across plate files, and never internally to any file, so that close doublets were preserved.  ``Best'' photometry was kept from matching objects, preferring two-band photometry to single-band, and preferring photometry with PSFs to that without.  3364 overlapping de-duplication exercises were done on the POSS-I plate files, 3269 on the POSS-II plate files, and 3171 on the UKST plate files. 

Each of the three surveys (POSS-I, POSS-II and UKST), having been fully calibrated and de-duplicated, was now assembled into whole sky data and partitioned into computing-friendly ``tiles'' of sky (of 10$^{\circ}$ RA x 9$^{\circ}$ DEC) with arcminute-wide margins added onto their edges to prevent edge effects.  Matching tiles across surveys were now overlain and their objects deduplicated one-to-one within a 4.75 arcsec radius (found via annulus counts), keeping ``best'' photometry as before.  Two-band POSS-I photometry was always retained as the first choice, followed by any other two-band photometry, else single-band.  Tycho-2 stars from the USNO-B were also added back in, matched within a 5.35 arcsec radius.  USNO-A UKST data was added where unmatched to any other.  The resultant APM/USNO-A/B data number 957\,205\,473 sources and represent an all-sky map of relatively uniform photometric depth, although less deep near the Galactic plane where high stellar density challenges instrument design limits.

The SDSS DR9 sweeps data was now merged onto this, identifying shared objects one-to-one within a 3.35 arcsec matching radius (found via annuli counts, refer Appendix Figure A2); the closer radius is because of the excellent astrometric accuracy of the SDSS data so that the astrometric errors are on the APM/USNO-A/B side only.  Lastly, the SDSS data from XDQSO \cite{XDQSO} was added where not already present, for completeness.  Because the uniformity of the APM/USNO-A/B photometry is potentially useful in research, the SDSS photometry overwrites it only where two-band APM/USNO-A/B photometry does not already exist.  However, the superior SDSS astrometry is always retained wherever matched.  This policy means that ASP presents 323\,126\,904 objects with SDSS astrometry, but of those only 206\,031\,717 bear SDSS photometry. 
                 
With merging completed, the sky tiles were cropped to their fiducial dimensions (thus freed of edge effects) and re-assembled into the 100 files presented on the website.  The completed ASP catalogue has 1\,163\,237\,190 sources each of which is flagged as to the provenance of its photometry and, separately, its astrometry.

\section{Methodology of the ASP catalogue's presentation}

Each ASP object is reported simply with its astrometry, magnitudes, PSFs and flags, but those data can be sourced from up to six places from four input catalogues.  To manage this, I endeavoured to select the best photometry and most recent-epoch astrometry for each object, via the following hierarchy of usage: 
\begin{itemize}
	\item For photometry, two-band photometry is always preferred to single-band.  Within that constraint, POSS-I is the top choice because of the wide spread between its red 6400\AA\ \textsl{E} and violet 4050\AA\ \textsl{O} bands, plus that its (\textsl{O-E}) colour is reliable due to both plates being taken on the same night.  Second preference goes to POSS-II and UKST (which both feature blue-green 4850\AA\ \textsl{Bj}) because they comprise a relatively uniform all-sky map which can be useful for research.  As a tie-breaker, APM photometry is preferred to USNO-A/B because of its recalibrated accuracy.  Lastly, the deeper SDSS photometry is ``clipped on'' where two-band photometry was not already present in the APM/USNO-A/B all-sky map.  In this way, researchers can retrieve the uniform APM/USNO-A/B sky map shown on Figure 3 by simply disregarding the SDSS photometric data.  
	\item For astrometry, the superior SDSS astrometry is always kept; its mean epoch is 2005 (observed 2000-2009).  Next, USNO-B projected their astrometry to epoch 2000 wherever two survey epochs were available, and ASP flags those accordingly.  Remaining objects have the mean epoch of the POSS-II, UKST or POSS-I astrometry with which they are flagged, being 1993 for POSS-II (observed 1985-2000), 1985 for UKST (observed 1978-1990), and 1955 for POSS-I (observed 1949-1958).  Note that the ``mean epochs'' used here have a typical uncertainty of $\pm$5 years to the true observational epoch of any datum.  Counts by catalogue \& survey are given in Table 1.  
\end{itemize}

\begin{figure*}[t] 
\includegraphics[scale=0.425, angle=0]{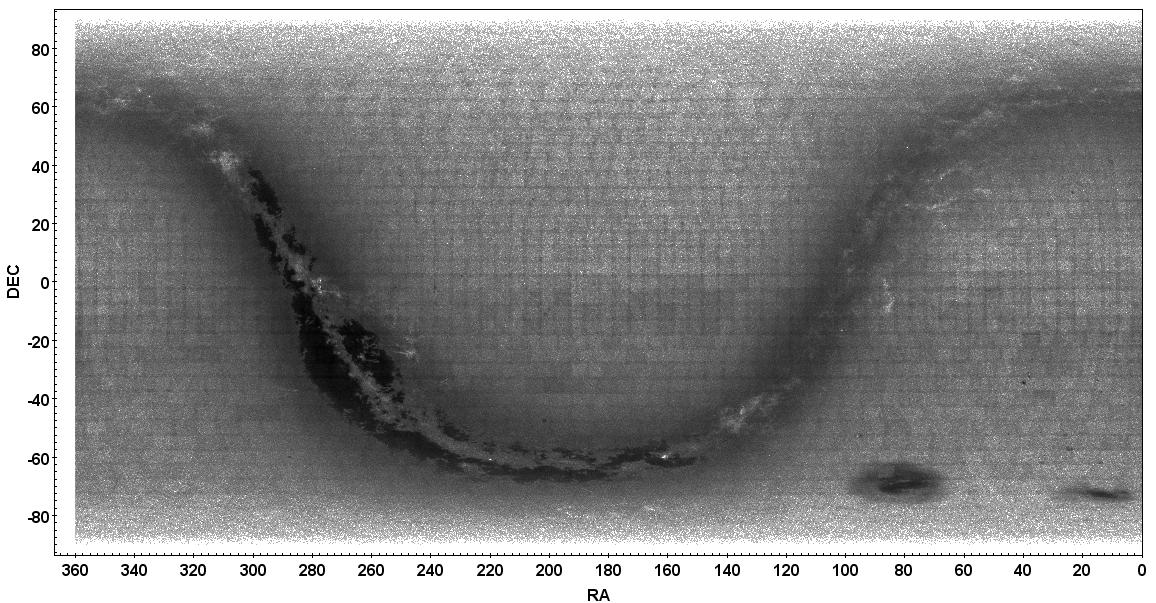} 
\caption{ASP sky coverage with SDSS photometric data removed.  Photographic plate lattices are seen throughout because plates' overlapping margins have denser data due to well-separated duplicates, unique data contributed by each plate, and USNO-B artefacts.}   
\end{figure*}

\begin{table}[ht]
\scriptsize	 
\caption{ASP Catalogue: Counts of sources by input catalogue \& survey for photometry \& astrometry}
\begin{tabular}{lrr}
Catalogue/Survey  &   \# Photometry  &    \# Astrometry \\
\hline
APM POSS-I        &    59\,767\,828  &         825\,037 \\
USNO-B POSS-I     &   306\,653\,894  &     20\,293\,648 \\
USNO-A POSS-I     &     9\,687\,099  &      4\,502\,407 \\
USNO-B POSS-II    &   118\,099\,510  &    136\,192\,015 \\
APM UKST          &    98\,324\,987  &     90\,395\,166 \\
USNO-B UKST       &   360\,089\,145  &    320\,554\,552 \\
USNO-A UKST       &     2\,032\,052  &      2\,032\,052 \\
SDSS DR9 Sweep    &   196\,934\,228  &    314\,029\,415 \\
SDSS DR8 XDQSO    &     9\,097\,489  &      9\,097\,489 \\
USNO-B Tycho-2    &     2\,550\,958  &                  \\
USNO-B Epoch 2000 &                  &    265\,315\,409 \\
\hline
Total             & 1\,163\,237\,190 & 1\,163\,237\,190 \\
\hline
\end{tabular}
\end{table}

The presented PSFs are usually from the photometry provenance but can come from either.  In very rare cases the red \& blue magnitudes are from different provenances because only single-band source data were available; in such cases the cited photometry provenance pertains to the blue magnitude.

The USNO-B gives a proper motion flag where stars were identified as moving; this is included in ASP where USNO-B's confidence of the proper motion is $>$90\%.  The nominal count is 157\,204\,744 objects so flagged, but USNO-B overreported proper motion incidence by $\approx$50x in a bid for completeness, so that flag should always be taken as indicative only and needing confirmation in individual cases.

ASP provides 25\,621\,092 objects with 1955-epoch POSS-I astrometry which can be moving stars matching to nearby later-epoch signatures, but are present for other reasons also: (1) objects close to bright stars or on galaxy disks which were not identified by later survey reductions, see Figure 4 for an example, (2) moving asteroids which are unmatchable, (3) variable objects which have faded, (4) POSS-I plate artefacts which would be invalid objects, and (5) POSS-II/UKST plate artefacts and edge effects which lose 2$^{nd}$-epoch stars, thus leaving those POSS-I objects unmatched.  Due to different epochs, moving stars can manifest 2 or even 3 times in ASP, each flagged with its own epoch, but I estimate the incidence of this is just $\approx$0.1\% of all objects and usually one of those matching sources will have the proper motion flag, see Figure 5 for an example.

\begin{figure}[ht] 
\includegraphics[scale=0.45, angle=0]{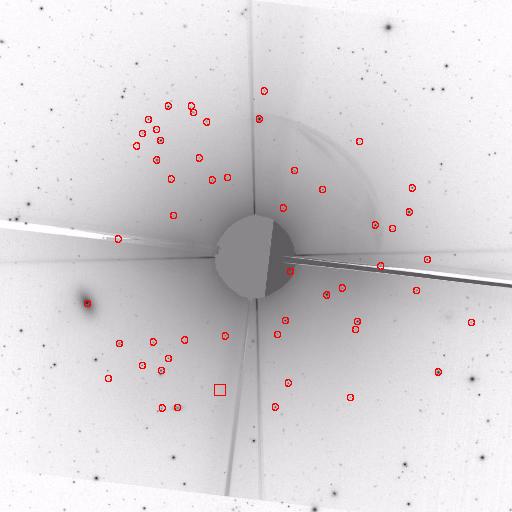} 
\caption{Sky around the star 31 Leo (J100754.3+095951) from a 12 arcmin-sq SDSS finding chart.  Red circled objects appear in ASP with epoch-1955 POSS-I astrometry from the APM \& USNO-A catalogues; they ring the masked central star because neither the SDSS nor USNO-B catalogues identifed them in the glare of the central star.  Objects farther out than 4.5 arcminutes' radius from the central star are reported with astrometry of later epochs from SDSS and USNO-B.  At J100757.9+095617, shown as a large red square at lower left, a stellar object of \textsl{R}=18.3 appears on the POSS-I plates but is not seen at later epochs.}  
\end{figure}

\begin{figure}[ht] 
\includegraphics[scale=0.50, angle=0]{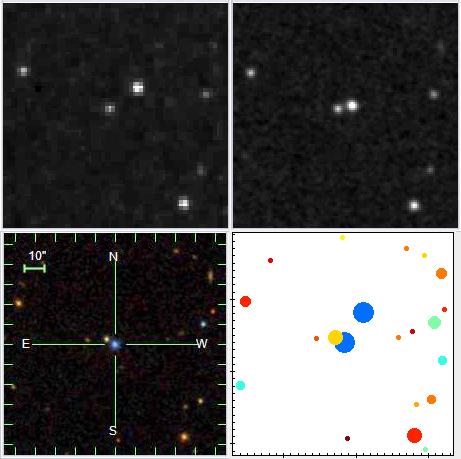} 
\caption{The white dwarf SDSS J090514.79+090426.2 across the epochs, all images are 2'x2' and centred on the SDSS position.  Upper left: POSS-I red plate from the 1950's; upper right: POSS-II red plate from the 1980's; lower left: SDSS chart from the 2000's.  Star shows proper motion with bearing of 147$^{\circ}$ E of N.  Lower right: this sky as represented in ASP: disk sizes represent \textsl{V} magnitude, red-blue colours using a rainbow palette.  The white dwarf star is present twice, once at the SDSS position at centre, and once at the POSS-I position which is flagged with epoch 1955.  ASP does flag the object at centre as showing proper motion -- flag taken from USNO-B.}      
\end{figure}

\section{Photometric Comparison of Catalogues}

The POSS-I and POSS-II/UKST surveys used red plates which replicated the standard Cousins \textsl{R} band, but their blue plates differed from  the standard Johnson \textsl{B} (4400\AA) band and from each other.  The SDSS \textsl{r} and \textsl{g} bands differ again.  ASP reports its magnitudes in those original survey photometries, but with just one survey per each object, identified with a flag.  The user may wish to standardize to \textsl{R} and \textsl{B} and to Johnson \textsl{V} for purposes of comparison, so some convenient literature-based transformations follow.  First, the standard colour transform from Minkowski \& Abell \shortcite{MA} is:     

\begin{center} \textsl{B - R = 1.6 ( B - V )} \end{center} 

\noindent Solving for \textsl{V}: 

\begin{equation} \textsl{V = {\large$\textstyle\frac{5}{8}$} R + {\large$\textstyle\frac{3}{8}$} B} \end{equation} 

\noindent For POSS-I \textsl{E} \& \textsl{O} (4050\AA) photometry, McMahon et al. \shortcite{MWHB} give this transform to Johnson \textsl{B}:

\begin{equation} \textsl{B = O - 0.12 ( O - E )} \end{equation} 

\noindent which, combining with (1) and taking \textsl{E} = \textsl{R}, gives:

\begin{equation} \textsl{V = 0.67 E + 0.33 O} \end{equation} 

\noindent For POSS-II and UKST photometry, Blair \& Gilmore \shortcite{BG} give this transform of \textsl{Bj} (4850\AA) with Johnson \textsl{B} and \textsl{V}: 

\begin{center} \textsl{Bj = B - 0.28$\pm$0.04 ( B - V )} \end{center} 

\noindent Similarly, Irwin/Demers/Kunkel \shortcite{IDK} give:

\begin{center} \textsl{Bj = B - 0.25$\pm$0.03 ( B - V )} \end{center} 

\noindent Those two transforms agree within their error margins, so taking their mean and combining with (1) gives:

\begin{equation} \textsl{V = 0.55 R + 0.45 Bj} \end{equation} 

\noindent For SDSS photometry, Lupton \shortcite{Lupton} gives:

\begin{center} \textsl{V = g - 0.5784 (g - r) - 0.0038} \end{center} 

All these transformations enable comparison of POSS-I, POSS-II/UKST and SDSS magnitudes by converting each to Johnson \textsl{V}.  The following transformations to Johnson \textsl{B} and Cousins \textsl{R} enables comparison of \textsl{(B - R)} colour: \\

For POSS-I data, \textsl{R = E}, and combining with (2):
\begin{center} \textsl{B - R = 0.88 ( O - E )} \end{center} 
 
For UKST \& POSS-II data, \textsl{R} equates to Cousins \textsl{R}:
\begin{center} \textsl{B - R = 1.32 ( Bj - R )} \end{center} 
 
For SDSS data, Lupton \shortcite{Lupton} gives:
\begin{center} \textsl{R = r - 0.1837 (g - r) - 0.0971} \end{center} 
\begin{center} \textsl{B = g + 0.3130 (g - r) + 0.2271} \end{center} 
\begin{center} thus: \textsl{B - R = 1.4967 (g - r) + 0.3242} \end{center}

The recent \textsl{Gaia} DR1 \cite{GAIA} presents optical data in a single broad (3300\AA - 10500\AA) \textsl{'G'} (for \textsl{'Gaia'}) band to their limit of \textsl{G} = 20.7.  It's interesting to compare the ASP optical depths to that limit, so I use the transformation from \textsl{V} \& \textsl{R} to \textsl{G} given by Jordi et al. \shortcite{GBAND}:

\begin{center} \textsl{G = V - 0.012 - 0.3502 ( V - R ) - 0.6105 ( V - R )$^{2}$ + 0.0852 ( V - R )$^{3}$} \end{center}
 
This transform enables mapping to \textsl{Gaia G} for all ASP data, since we already have transforms to \textsl{V} \& \textsl{R}.  

These transforms enable us to compare the plate depths of the catalogue/surveys included in ASP.  APM gives \textsl{R/Bj} or \textsl{E/O} plate depths for its 959 plates, and their mean depths are shown in Table 2 along with USNO-A/B mean plate depths derived from magnitude profiles.  SDSS ``depth'' is a nebulous concept but I include it in Table 2 by taking the SDSS data frequency maxima of \textsl{r=22.50} and \textsl{g=22.98} in the SDSS data as surrogates for depth; perhaps deeper values would be justified.  The mappings to \textsl{V} and \textsl{G} show:
\begin{enumerate}
  \item SDSS coverage is $\approx$2 magnitudes deeper than other coverage.  Runner-up is the APM Southern UKST coverage which at average plate depth of \textsl{V}=21.68 is also impressively deep.
	\item APM data is deeper than USNO-A/B because of its deep blue band coverages which range 0.6 to 1.0 magnitude deeper than that of USNO-A/B.  This may have been consequential to APM's use of contrast-enhanced glass plate copies which removed all dynamic range but revealed the faintest objects, which appears to have been particularly effective with blue plates.   
	\item USNO-A/B plate depths away from the Galactic plane (i.e., onto the APM footprint shown in Figure 2) reach about a magnitude deeper than those onto the Galactic plane.  Also, very dense places such as the Galactic Bulge and the LMC \& SMC have shallower (and unreliable) coverage, due to operational limits of the plate measuring machine.   
	\item SDSS coverage is $\approx$2 magnitudes deeper than the \textsl{Gaia} limit of \textit{G}=20.7.  South sky UKST coverage away from the Galactic plane reaches \textsl{G} = 21.0, also deeper than the \textsl{Gaia} limit.   
\end{enumerate}

\begin{table}[ht]
\scriptsize	 
\caption{Catalogue-Survey mean plate depths}
\begin{tabular}{@{\hspace{0pt}}l@{\hspace{4pt}}c@{\hspace{8pt}}l@{\hspace{8pt}}c@{\hspace{8pt}}c}
Catalogue       & \textsl{R$_{C}$}           & blue band &             & \textsl{Gaia} \\
\& Survey       & depth & \& depth  & \textsl{V} & \textsl{G}    \\
\hline
SDSS all footprint (estimated)         & 22.31 & \textsl{g} = 22.98  & 22.70 & 22.47 \\ 
\hline
\multicolumn{5}{@{\hspace{0pt}}l}{\textit{-- plates away from Galaxy, i.e., in APM footprint (Fig. 2) --}} \\
USNO-B POSS-II ($\delta>-3.2^{\circ}$) & 20.38 & \textsl{Bj} = 21.55 & 20.90 & 20.55 \\
APM POSS-I ($\delta>-3.2^{\circ}$)     & 20.16 & \textsl{O} = 21.97  & 20.76 & 20.34 \\
USNO-B POSS-I ($\delta>-33^{\circ}$)   & 20.29 & \textsl{O} = 21.35  & 20.64 & 20.43 \\
USNO-A POSS-I ($\delta>-33^{\circ}$)   & 19.90 & \textsl{O} = 21.30  & 20.36 & 20.07 \\
APM UKST ($\delta<+3.2^{\circ}$)       & 20.86 & \textsl{Bj} = 22.70 & 21.68 & 21.02 \\
USNO-B UKST ($\delta<-3.2^{\circ}$)    & 20.94 & \textsl{Bj} = 21.68 & 21.27 & 21.08 \\
USNO-A UKST ($\delta<-33^{\circ}$)     & 20.90 & \textsl{Bj} = 22.10 & 21.43 & 21.07 \\
\hline
\multicolumn{5}{@{\hspace{0pt}}l}{\textit{-- plates onto the Galactic plane --}} \\
USNO-B POSS-II ($\delta>-3.2^{\circ}$) & 19.49 & \textsl{Bj} = 20.85 & 20.10 & 19.67 \\
USNO-B POSS-I ($\delta>-33^{\circ}$)   & 19.08 & \textsl{O} = 20.73  & 19.62 & 19.25 \\
USNO-A POSS-I ($\delta>-33^{\circ}$)   & 19.15 & \textsl{O} = 20.95  & 19.74 & 19.33 \\
USNO-B UKST ($\delta<-3.2^{\circ}$)    & 19.53 & \textsl{Bj} = 20.70 & 20.05 & 19.70 \\
USNO-A UKST ($\delta<-33^{\circ}$)     & 19.40 & \textsl{Bj} = 20.80 & 20.02 & 19.58 \\
\hline
\multicolumn{5}{@{\hspace{0pt}}l}{Note: Only two-band stellar objects were used in this analysis.} \\ 
\multicolumn{5}{@{\hspace{0pt}}l}{Calibration of these catalogue-surveys is discussed in Section 3.} \\
\end{tabular}
\end{table}

\section{The Sky as Presented by ASP}

The ASP data is found to have astrometric accuracy \textsl{onto the SDSS epoch} of 90.96\% within 1 arcsecond and 96.65\% within 2 arcseconds.  The excellent SDSS astrometry is used as the benchmark for this calculation.  In ASP, 117\,095\,187 APM/USNO-A/B objects match to SDSS objects within the matching radius of 3.35 arcseconds.  Of those, 87.49\% match within 1 arcsec and 95.37\% match within 2 arcsec; this is taken as representative of all 840\,110\,286 ASP objects without SDSS astrometry.  The 323\,126\,904 ASP objects with SDSS astrometry are taken as perfectly positioned.  Combining these two groups, the overall astrometric accuracy works out to that stated above.  For stellar-PSF objects only, the match rates over 47\,830\,533 shared objects are 81.64\% matching within 0.5 arcsec, 92.99\% within 1 arcsec, and 97.48\% within 2 arcsec, which can be taken as the practical astrometric accuracy of ASP sky away from the SDSS footprint.  

The ASP data is a useful listing of the photographed sky, but its survey-based coverage is inhomogeneous on both large and small scales, and shows processing artefacts.  To confer some overview, some ASP sky tiles are shown as density maps, refer Figures 6 to 10 with discussions thereon.  Individual photographic plates are clearly seen because plate overlaps are denser due to unique data contributed by each plate, plus some well-offset duplicates and artefacts.

\begin{figure}[ht] 
\includegraphics[scale=0.325, angle=0]{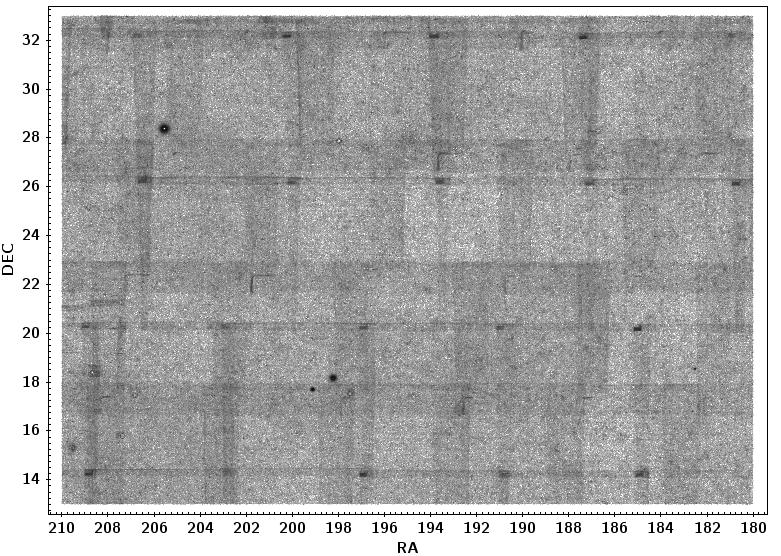} 
\caption{Northern Galactic Cap sky density tile with SDSS objects removed, darker is denser.  Two photographic plate grids are seen:  POSS-I plates with thin overlaps are arrayed at 6$^{\circ}$ intervals and show a small USNO-B artefact of spurious data at the upper left corner of each plate, probably from a rectangular label.  POSS-II plates constitute another lattice and are seen throughout with broad overlaps arrayed at 5$^{\circ}$ intervals.  Overlap zones are denser because of unique data contributed by each plate, plus some far-offset duplicates.}   
\end{figure}

\begin{figure}[ht] 
\includegraphics[scale=0.325, angle=0]{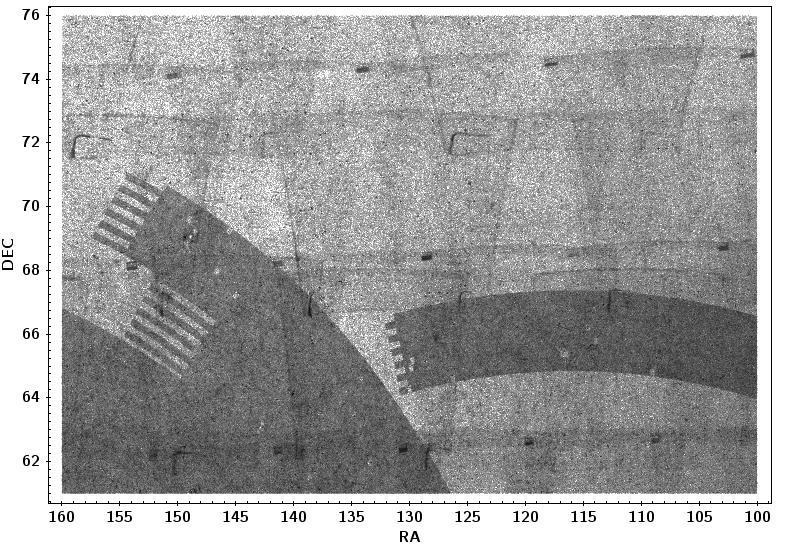} 
\caption{Northern high-latitude sky density.  The edge of the main SDSS footprint and an SDSS extension arc are seen across the bottom half.  APM \& USNO-B POSS-I plates (with thin margins and small USNO-B label artefact at their upper left corners) and USNO-B POSS-II plates (with broad margins and label edge artefact at their lower right corners) are seen to overlap differentially as they near the pole.}  
\end{figure}

\begin{figure}[ht] 
\includegraphics[scale=0.38, angle=0]{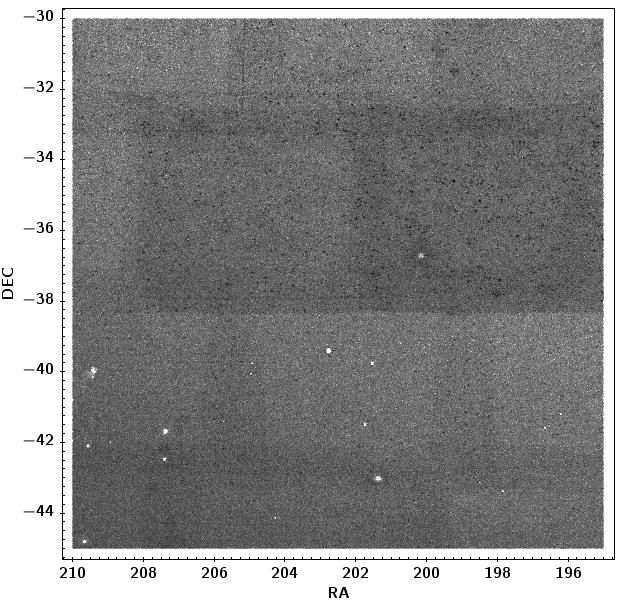} 
\caption{Southern sky density, comprised primarily of UKST plates, is dense and comparatively uniform.  Denser APM data darkens the upper half of this 15$^{\circ}$x15$^{\circ}$ tile to $\delta$=$-38.2^{\circ}$; this shows the southward limit of the APM data in the North Galactic Cap.  Also, $\delta$=$-33.2^{\circ}$ is the southward limit of supplementing POSS-I plates from USNO-A/B.}
\end{figure}

\begin{figure}[ht] 
\includegraphics[scale=0.38, angle=0]{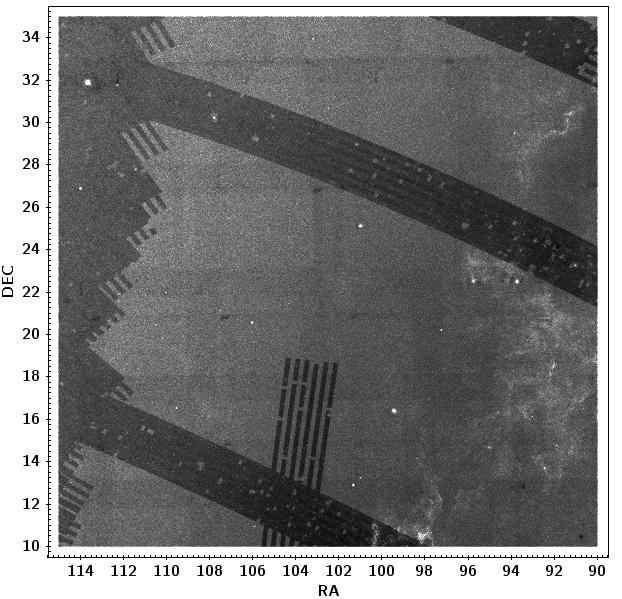} 
\caption{Sky density crossing the Galactic anticentre.  The main SDSS footprint at left shows a ragged edge; Galactic dust filaments are seen at right, with SDSS extensions crossing over the Galactic plane.  The background photographic plate pattern shows a mix of USNO-B POSS-II plates with broad margins, and USNO-A/B POSS-I plates with thin margins and small dark artefact at the corners.  The background density increases (darkens) from left to right on the Galactic approaches.} 
\end{figure}

\begin{figure}[ht] 
\includegraphics[scale=0.32, angle=0]{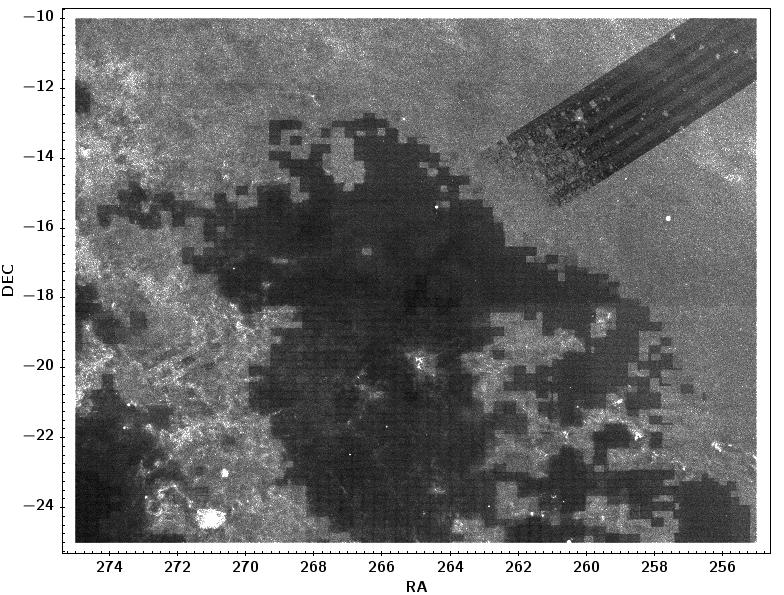} 
\caption{Sky density toward the Galactic centre, North end of the Bulge at centre.  An SDSS extension thrusts toward the Bulge which however was not reached; the Bulge is so dense with stars that instrument design limits typically get exceeded as was certainly expected in this test run.  The USNO-B coverage of the Bulge was similarly impaired: the small dark rectangles there show 20'x16' CCD footprints with saturated sky values and consequently unreliable data.  Note traces of the continuing SDSS extension at lower left where it reached the Galactic dust lane.} 
\end{figure}

Along those lines, a few caveats to keep in mind:  
\begin{enumerate}
	\item Fuzzy (galaxy-like) objects from the APM or USNO-B usually display magnitudes which are too bright.  This is because the magnitudes were calculated from the plate image size using a stellar (not fuzzy) PSF model.  For \textsl{R}$>$16, the mean conversion to true magnitude is \textsl{R$_{\textit{true}}$ = $\textstyle\frac{2}{3}$(11+R)} but the scatter is too great to apply this individually.  Divergence increases non-linearly for \textsl{R}$<$16 and can be many magnitudes too bright for \textsl{R}$<$10.  	
	\item USNO-A/B had difficulty setting sky levels in high-density places like the Magellanic Clouds and Galactic bulge, so the outcome there was spurious data interwoven with valid data.  Use with great caution in those places.
	\item USNO-A/B plates show label-caused artefacts.  Their POSS-I plates show a small rectangular label in the upper left corner which manifests as spurious \textsl{r$\approx$20.5 b$\approx$21.1} data there.  The USNO-B POSS-II plates have an opaque label on the lower right corner which is without data but is fully covered by neighbouring plates; however, an edge-shaped artefact resulted because objects didn't get de-duplicated across the edge there.  They can be seen in Figures 6 and 7. 
	\item USNO-B included objects which appeared at least twice in their 5 bands of data (being first-epoch red \& blue and second-epoch red, blue, and infrared).  Thus, since ASP reports just 2 bands of data, many USNO-B objects appear as just single-band and those are found to have reduced reliability.  Single-band objects on the SDSS footprint are not matched to SDSS data (else they would be presented with two-band SDSS photometry) and so are often false.  Single-band objects comprise 12.89\% of all objects in ASP and when selecting one it would be best to confirm its existence on a DSS\footnote{Digitized Sky Survey, http://archive.stsci.edu/cgi-bin/dss\_form} image or SDSS finding chart.
	\item Some well-offset duplicates remain in the data, including cases where a fuzzy-PSF object conceals a point source which was nontheless contributed by another source catalogue.  De-duplication processing can also (rarely) copy photometric attributes across close neighbours.  And duplicates from different epochs can be present, refer Figure 5.  The user should be alert to the possible presence of such duplicates, and check with DSS or SDSS as necessary.  
\end{enumerate}

\section{Catalogue Layout}

The ASP catalogue uses an efficient data design to minimize its size.  Part of this design is that RA and DEC are presented as offsets from stated zero points.  The RA zero point is given by a 1-byte field which identifes the tenth of a RA hour, e.g., ``117'' $\Rightarrow$ 11.7 RA hours = 175.5 RA degrees.  The DEC zero point is given by the file name x 1.8$^{\circ}$, e.g. ``N17'' $\Rightarrow$ 17 x 1.8$^{\circ}$ = 30.6$^{\circ}$.  This design allows RA and DEC to be presented as simple 2-byte fields which give the offset from its zero point in units of 0.1 arcsecond.  

Accordingly, this ASP catalogue presents each record as a 10-byte binary row, with 1 byte for the RA zero point, 2 bytes each for the RA \& DEC offsets, 2 bytes each for red and blue photometry, and 1 byte for flags.  RA \& DEC fields use an unsigned integer binary little-endian format as are used in PCs nowadays; alternatively they can be input as signed integers, and where the resultant value is negative, add 65536.  

The ASP catalogue comes in 100 files, each of which holds all RA for a 1.8$^{\circ}$-wide (in DEC) ring of sky.  These 100 rings thus stack the sky from pole to pole.  The Southern files (S00.asp - S49.asp) are written just like Northern files (N00.asp - N49.asp) with all declinations written as positive numbers only.  The user should process them identically to the Northern files, treating South as North, and convert the declinations to negative at the conclusion by simply affixing the ``-''.   

The ReadMe gives a thorough discussion of all the fields, and includes some BASIC and Python code which performs the processing.  It is available on the file download web page.  

The All-Sky Portable (ASP) catalogue is available for download as 15 zipped files of 5+ ASP files each\footnote{The ASP files are available on the PASA datastore at http://dx.doi.org/10.4225/50/5807fbc12595f .  The author's ASP home page is at http://quasars.org/asp.html .}, with total size of 9Gb zipped, 11Gb unzipped.  No FITS files are provided due to the large size of the data.  There is no on-line query client; ASP is provided only as a bulk catalogue for download.

\section{Miscellaneous Notes}
 
There are 102\,048 ``inferred'' optical magnitudes in the ASP data.  They are legacy data from Flesch \& Hardcastle \shortcite{QORG} which joined USNO-A2.0 point sources to APM isophotal ellipses to find cases where two point sources were required to generate the APM ellipse but only one was reported -- therefore the missing point source could be astrometrically calculated, i.e., inferred.  Since then, the addition of USNO-B data has filled in most of those missing data.  In ASP, they are used only to fill out the photometry for single-band sources, thus showing that the object is seen in the other band also.  Inferred magnitudes have an expected error of 1 magnitude and can be greater.  

Variability greater than 1 magnitude was evaluated across overlapping plates throughout the assembly of this catalogue.  Such variability is flagged in both red \& blue bands for 3.28\% of all objects.  However, it should be treated as indicative only and needing confirmation in individual cases.

\section{Conclusion}

This ``All-Sky Portable'' (ASP) optical catalogue is presented with 1\,163\,237\,190 optical sources over the whole sky, taken from earlier optical catalogues produced in the years 1996-2003 and which are no longer available in bulk, plus SDSS data.  The data is in a binary format of tenth-of-arcsecond precision astrometry, hundredth-of-magnitude precision red-blue photometry with stellar-fuzzy PSFs, and flags for proper motion, variability, epoch, and provenance of each of photometry and astrometry.  The catalogue is downloadable at a total zipped size of 9Gb.                          

\begin{acknowledgements}
Great thanks to Dave Monet (USNO) and Mike Irwin (APM) for permission to use their optical catalogues in this work, and for helpful advice.  Also, much gratitude to Adam Myers for provision of the SDSS sweeps data which has made ASP complete.  An extra shout-out to Dave Monet for heavy lifting: 10\,000 survey plates scanned, amazing.
  
This work was not funded.
  
\end{acknowledgements}

%\clearpage
\appendix{\textbf{\large{APPENDIX}}} 
\section{Use of Annuli to Determine Matching Rules}

A substantive task in producing ASP was to avoid the inclusion of duplicate objects.  The three main sources of duplicates were input catalogues presenting the same sky from the same survey, or the same sky from different surveys, or adjacent survey plates with margins overlapping each other.  These were matched using a one-to-one matching algorithm which matches astrometrically closest objects and removes those objects from their respective pools, then repeats iteratively to completion.  But the overriding question is what maximum matching radius (i.e., offset) to use to identify those duplicates, because matches at large offsets are more likely to be discrete objects.  Such true objects constitute a background which we wish to preserve, so for large processing we need a statistical approach to discern the duplicates from the true background.  This involves counting the number of matches for each binned matching radius and using those counts to understand the data.  However, doing so baldly isn't good enough, as is shown in Figure A1.

\begin{figure}[ht] 
\includegraphics[scale=0.38, angle=0]{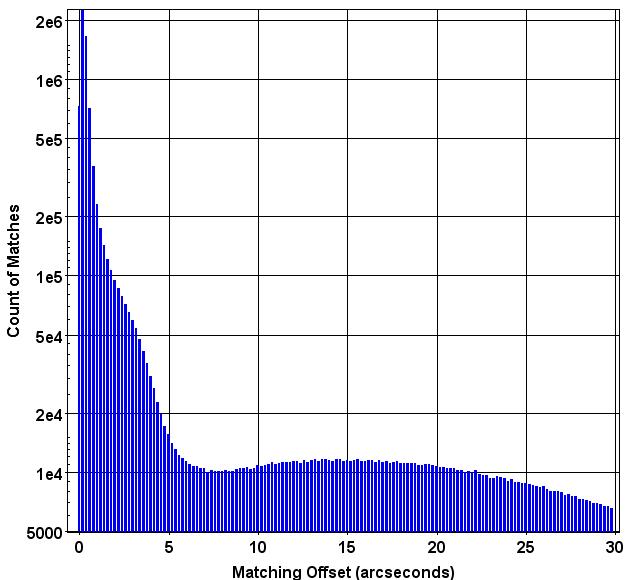} 
\caption{One-to-one matching of SDSS data to APM/USNO-B combined data, binned to annuli of 1/5$^{th}$ arcsecond width.  The vertical scale is logarithmic to improve visibility at low counts, n = 8\,733\,411 matches for this chart.}
\end{figure}

Figure A1 isn't helpful: for large matching offsets the count profile is non-linear and not even monotonic, so the background of true objects is not seen.  The solution is two-fold: (1) use many-to-one matching instead of one-to-one, and (2) instead of counts of matches, present the density of matches, where the area used is that of the 0.1-arcsec-wide annulus on the sky for each matching offset (i.e., radius) bin which works out simply to:    

\begin{center} Annulus area = {\large$\textstyle\frac{\pi}{5}$} R (in sq. arcsec) \end{center} 

Using this, the background of true objects for the SDSS-to-APM/USNO-B matching exercise is revealed as shown in Figure A2 (zoom in to see well).  

\begin{figure}[ht] 
\includegraphics[scale=0.38, angle=0]{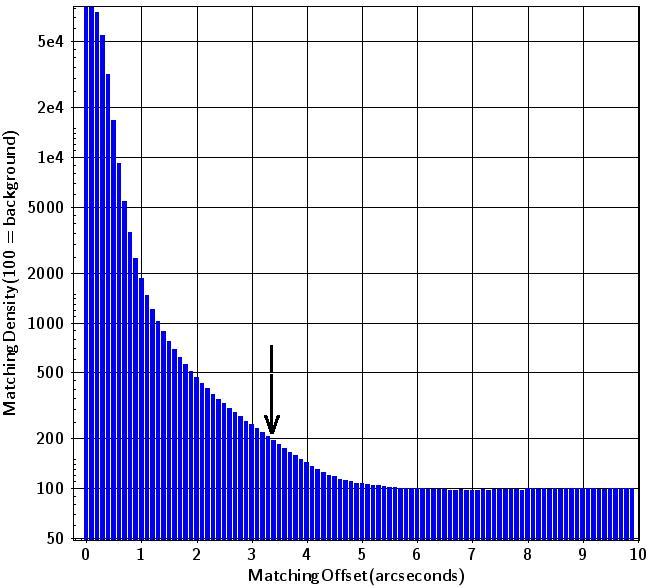} 
\caption{Many-to-one matching of SDSS data to APM/USNO-B combined data, binned to annuli of 1/10$^{th}$ arcsecond width.  The vertical scale is logarithmic to improve visibility at low counts; n = 11\,567\,232 matches over 1170 sq degrees for this chart.  The background density is normalized to y=100; therefore, the chart area below the line y=100 represents the true background of discrete objects, with chart area above y=100 representing duplicates.  The arrow shows where the profile crosses the y=200 line -- there, at x=3.35, the count of duplicates equals the background count of discrete objects.}
\end{figure}

SDSS photometry is deeper than APM/USNO-B with twice the optical density on the sky; because of this, true SDSS doublets are plentiful.  The priority here is to preserve true SDSS doublets equally with removing the duplicates, so the crossover point is where their frequency is equal on the sky, i.e., at y=200 on Figure A2.  This line is crossed between the bins of x=3.3 and 3.4 (marked with the arrow), so the matching radius used in ASP processing was 3.35 arcseconds, although the final matching processing was one-to-one.

Therefore for each matching task it was first necessary to do a many-to-one matching exercise to determine the true background and decide the limiting matching radius, and then to apply that matching radius in a one-to-one matching process which performed the actual de-duplication.  The one-to-one matching preserves close doublets better than many-to-one does, thus enabling farther out matching, but I could not know how much farther, so I left it where it was.  

Additional bar charts below show, for some of the matching tasks encountered in ASP processing, the count density profiles, the matching radius decided on, and any additional considerations described in the text thereon.  The charts are best viewed at a high zoom.

\begin{figure}[ht] 
\includegraphics[scale=0.38, angle=0]{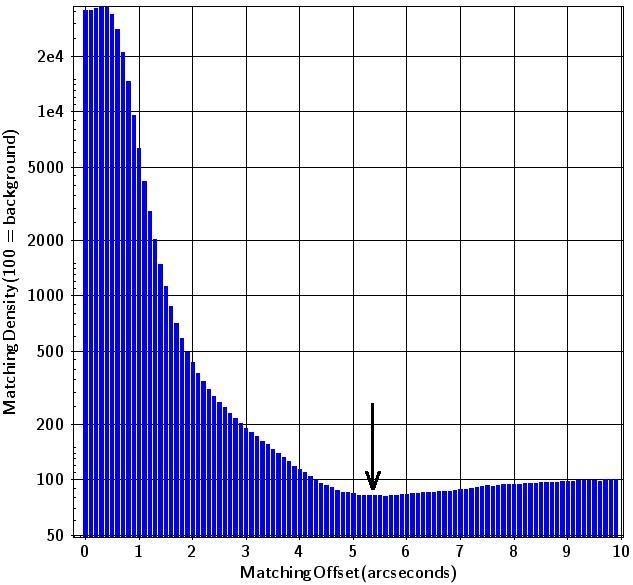} 
\caption{Many-to-one matching of APM to USNO-B data in the UKST survey area, binned to annuli of 1/10$^{th}$ arcsecond width with logarithmic vertical scale; n = 11\,731\,786 matches from 50 Schmidt plates for this chart.  The profile is hollowed out below background from x=4.4 to 9.  This appears to be because APM reports objects closer to bright stars than does USNO-B, see e.g., Figure 4; in matching them, this simulates an internal edge effect within USNO-B data because true objects are missing across the edges of the effective holes, thus lowering matching rates below background.  The matching radius was set at the inflection point of 5.35 arcseconds (indicated by the arrow) in tandem with the radius used for the POSS-I data which has a similar profile.}
\end{figure}
                 
\begin{figure}[ht] 
\includegraphics[scale=0.38, angle=0]{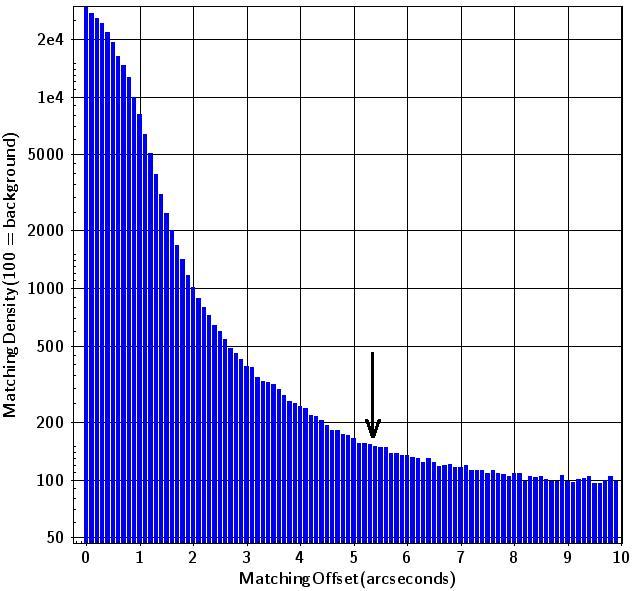} 
\caption{Many-to-one matching of full POSS-I Schmidt plate data with neighbouring plates with narrow overlaps (see depiction on Figure 6), binned to annuli of 1/10th arcsecond width; n = 436\,832 matches over 83 two-plate overlaps.  The priority here was to remove duplicates, more importantly than retaining close doublets.  Therefore the desired matching radius was farther out than the (y=200) crossing point which here is at x=4.45.  I selected 5.35 arcseconds simply to be consistent with other matchings, shown by the arrow. The unremoved duplicates in the area x$>$5.35 and y$>$100 comprise a residue which can reach large offsets because systematic offsets at plate edges can point oppositely to that of overlapping plates.}
\end{figure}

\begin{figure}[ht] 
\includegraphics[scale=0.38, angle=0]{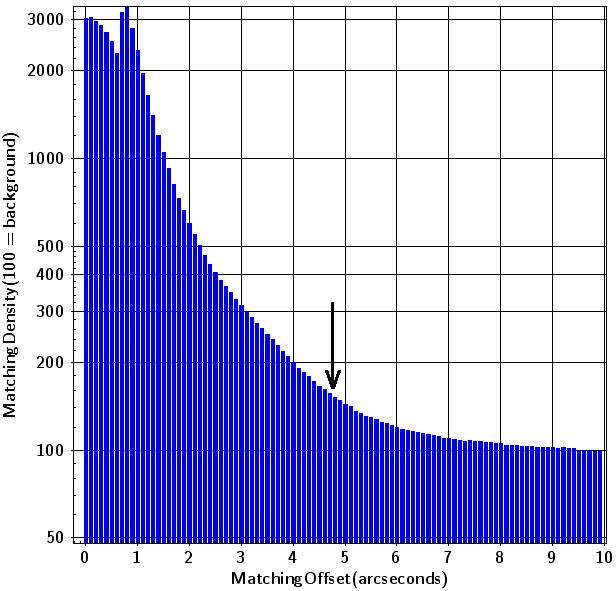} 
\caption{Many-to-one matching of multi-catalogue POSS-I data against attenuated$^{*}$ USNO-B POSS-II data, binned to annuli of 1/10th arcsecond width; n = 10\,880\,162 matches over about 8000 sq degrees of sky.  The priority here was to remove duplicates, but there was increased scope for retaining true objects from each survey which had no counterpart in the other survey.  There was no absolute answer in choosing the matching radius, so I chose a midpoint between the usual value of 5.35 and the y=200 crossing at x=4.0; thus, I decided on 4.75 arcseconds as the matching radius, indicated by the arrow.  The double peak at x=0, x=0.7 is because USNO-B data granularity is 0.7 arcsec, so matches closer than that are to APM or USNO-A data, and the peak at x=0.7 arcsec is largely from adjacent USNO-B data.} 
\scriptsize{* Attenuated because ASP had already selected POSS-I photometry over POSS-II photometry from the USNO-B data which had both, thus reducing the POSS-II population but increasing their uniqueness as is seen here in the relatively high background of discrete objects.}
\end{figure}

\end{document}